\documentstyle[12pt,aaspp4,eqsecnum,flushrt]{article}
\input{epsf.sty}
\lefthead{ROBERGE}
\righthead{Adiabatic Approximation}
\newcommand{\be}{\begin{equation}}
\newcommand{\ee}{\end{equation}}

\newcommand{\vecJ}{\mbox{\boldmath$J$}}
\newcommand{\veca}{\mbox{\boldmath$a$}}      
\newcommand{\vecB}{\mbox{\boldmath$B$}}      
\newcommand{\tBar}{\mbox{$t_{\rm Bar}$}}
\newcommand{\tBarJt}{\mbox{$t_{\rm Bar}\left(\Jtherm\right)$}}
\newcommand{\tgas}{\mbox{$t_{\rm gas}$}}

\newcommand{\Ipar}{\mbox{$I_{\parallel}$}}
\newcommand{\Iper}{\mbox{$I_{\perp}$}}
\newcommand{\Ts}{\mbox{$T_{\rm s}$}}
\newcommand{\Tg}{\mbox{$T_{\rm g}$}}
\newcommand{\TsTg}{\mbox{$\Ts/\Tg$}}
\newcommand{\rhos}{\mbox{$\rho_{\rm s}$}}

\newcommand{\vth}{\mbox{$v_{\rm th}$}}
\newcommand{\Gpar}{\mbox{$\Gamma_{\parallel}$}}
\newcommand{\Gper}{\mbox{$\Gamma_{\perp}$}}
\newcommand{\Lint}{\mbox{${\cal L}_{\rm int}$}}
\newcommand{\vint}{\mbox{$\theta$}}
\newcommand{\Lintv}{\mbox{$\Lint\left(\vint\,|\,\vext\right)$}}
\newcommand{\Lext}{\mbox{${\cal L}_{\rm ext}$}}
\newcommand{\vext}{\mbox{$x$}}
\newcommand{\xset}{\mbox{$\left(x_1,x_2,x_3\right)$}}
\newcommand{\Lextv}{\mbox{$\Lext\left(\vext\,|\,\vint\right)$}}
\newcommand{\fjoint}{\mbox{$f$}}

\newcommand{\At}{\mbox{$A_{\theta}$}}
\newcommand{\fint}{\mbox{$f_{\rm int}$}}
\newcommand{\fintv}{\mbox{$\fint\left(\vint\,|\,J\right)$}}
\newcommand{\fext}{\mbox{$f_{\rm ext}$}}
\newcommand{\fextv}{\mbox{$\fext\left(\vext\right)$}}
\newcommand{\Atv}{\mbox{$\At\left(J,\theta\right)$}}

\newcommand{\Bttv}{\mbox{$B_{\theta\theta}(J,\theta)$}}
\newcommand{\Abm}{\mbox{$\bar{A}_m$}}
\newcommand{\Abmv}{\mbox{$\Abm\left(\vext\right)$}}
\newcommand{\AbJ}{\mbox{$\bar{A}_J$}}

\newcommand{\Bbmn}{\mbox{$\bar{B}_{mn}$}}
\newcommand{\Bbmnv}{\mbox{$\Bbmn\left(\vext\right)$}}
\newcommand{\BbJJ}{\mbox{$\bar{B}_{JJ}$}}

\newcommand{\snt}{\mbox{$\sin\theta$}}

\newcommand{\Am}{\mbox{$A_m$}}
\newcommand{\Amv}{\mbox{$A_m\left(\theta,x\right)$}}
\newcommand{\Bmn}{\mbox{$B_{mn}$}}
\newcommand{\Bmnv}{\mbox{$B_{mn}\left(\theta,x\right)$}}
\newcommand{\eps}{\mbox{$\epsilon$}}
\newcommand{\Order}{\mbox{${\cal O}$}}
\newcommand{\Ordeps}{\mbox{${\cal O}(\epsilon)$}}
\newcommand{\vecx}{\mbox{\boldmath$x$}}      
\newcommand{\vecy}{\mbox{\boldmath$y$}}      
\newcommand{\vecz}{\mbox{\boldmath$z$}}      
\newcommand{\bfxh}{\mbox{$\hat{\vecx}^{b}$}}
\newcommand{\bfyh}{\mbox{$\hat{\vecy}^{b}$}}
\newcommand{\bfzh}{\mbox{$\hat{\vecz}^{b}$}}
\newcommand{\bfbasis}{\mbox{$\left(\bfxh,\bfyh,\bfzh\right)$}}
\newcommand{\Jtherm}{\mbox{$J_{\rm therm}$}}

\newcommand{\Jgpolar}{\mbox{$\left(J,\beta,\phi\right)$}}
\newcommand{\Jgcart}{\mbox{$\left(J_x,J_y,J_z\right)$}}

\newcommand{\Bper}{\mbox{$B_{\perp}$}}
\newcommand{\Bpar}{\mbox{$B_{\parallel}$}}

\newcommand{\QX}{\mbox{$Q_X$}}

\newcommand{\QXMax}{\mbox{$Q_{X,{\rm Max}}$}}

\newcommand{\Tfac}{\mbox{$\left(1+\Ts/\Tg\right)$}}

\newcommand{\gfxh}{\mbox{$\hat{\vecx}$}}
\newcommand{\gfyh}{\mbox{$\hat{\vecy}$}}
\newcommand{\gfzh}{\mbox{$\hat{\vecz}$}}
\newcommand{\gfbasis}{\mbox{$\left(\,\gfxh,\gfyh,\gfzh\,\right)$}}
\begin{document}
\title{An adiabatic approximation for grain alignment theory}

\author{W. G. Roberge}
\affil{Dept.\ of Physics, Applied Physics \& Astronomy}
\affil{Rensselaer Polytechnic Institute, Troy, NY 12180, USA, roberw@rpi.edu}

\author{To appear in {\it Monthly Notices of the Royal Astronomical Society}}

\begin{abstract}
The alignment of interstellar dust grains is described by the
joint distribution function for certain ``internal''
and ``external'' variables, where
the former describe the orientation of a grain's axes with respect
to its angular momentum, \vecJ, and the latter describe the
orientation of \vecJ\ relative to the interstellar magnetic field.
I show how the large disparity between the dynamical timescales
of the internal and external variables--- which is typically
2--3 orders of magnitude--- can be exploited to 
greatly simplify calculations of the required distribution.
The method is based on an ``adiabatic approximation'' which closely
resembles the Born-Oppenheimer approximation in quantum mechanics.
The adiabatic approximation prescribes an analytic distribution function
for the ``fast'' dynamical variables and a simplified Fokker-Planck
equation for the ``slow'' variables which can be solved straightforwardly
using various techniques.
These solutions are accurate to \Ordeps, where $\eps$
is the ratio of the fast and slow dynamical timescales.
As a simple illustration of the method, I derive an analytic
solution for the joint distribution established when Barnett relaxation
acts in concert with gas damping.
The statistics of the analytic solution agree with the results of
laborious numerical calculations which do not exploit the adiabatic
approximation.
\end{abstract}

\keywords{dust --- polarization --- ISM: magnetic fields}
\setcounter{footnote}{0}

%
%%%%%%%%%%%%%%%%%%%%%%%%%%%%%%%%%%%%%%%%%%%%%%%%%%%%%%%%%%%%%%%%%%%%%%%%%%%%
% 1. Introduction
%%%%%%%%%%%%%%%%%%%%%%%%%%%%%%%%%%%%%%%%%%%%%%%%%%%%%%%%%%%%%%%%%%%%%%%%%%%%
%
\section{Introduction}

Rapid advances in polarimetry and related observations have provided
tantalizing clues about the mysterious mechanism which aligns the
interstellar dust grains (for recent reviews,
see Aitken 1996; Hildebrand 1996; Jones 1996; Goodman 1996; and Whittet 1996;
see also
Arce et al.\ 1997; Chrysostomou et al.\ 1995; Gerakines et al.\ 1995;
Goodman \& Whittet 1995; Goodman et al.\ 1995; and Martin 1995).
The renewed interest in interstellar polarization has
stimulated theoretical activity on the longstanding
``grain alignment problem.''
The classical alignment mechanisms of Davis \& Greenstein
(1951), Gold (1952) and Purcell (1979 [P79]) have
been reinvestigated and modified to include effects that were
unknown or poorly understood at the time of the original
investigations (Roberge, DeGraff \& Flaherty 1993 [RDGF93];
Lazarian 1994 [L94], 1995a,b,c; 1997a,b;
Lazarian \& Roberge 1997; Lazarian \& Draine 1997).
It has been demonstrated that Gold-type alignment
can occur if the grains rotate with superthermal kinetic
energies (Lazarian 1995d) and that Gold's mechanism
operates efficiently in nonlinear Alfv\'{e}n
waves (Roberge \& Hanany 1990; L94) and regions undergoing
supersonic ambipolar diffusion (Roberge, Hanany \& Messinger 1995;
Desch \& Roberge 1997).
The possibility that anistotropic radiation fields can spin up the grains
(Dolginov \& Mytrophanov 1976; Dolginov \& Silant'ev 1976) has
been confirmed in realistic simulations of the radiative torques
(Draine \& Weingartner 1996) and it has
been demonstrated (Draine \& Weingartner 1997) that such
torques may play a direct role in the alignment process.

This is the second in a series of papers, the collective purpose
of which is to provide quantitative predictions on the efficiencies
of different grain alignment mechanisms.
I have shown elsewhere (Roberge 1996) that, although
it is generally impossible to infer the alignment efficiency
directly from observations, calculations of this type
can be used in conjunction with observations to test any
alignment mechanism unambiguously.
The first paper in the series (RDGF93) described a mathematical
technique which calculates the alignment by solving the Langevin
equation for Brownian rotation.
However, RDGF93 assumed that a grain's angular momentum
is aligned perfectly with its axis of largest rotational
inertia (its ``major axis of inertia'')
due to the rapid dissipation of rotational energy by the Barnett effect (P79).
Unfortunately, this is not an accurate assumption if the grains
rotate with thermal kinetic energies:
In an important paper, L94 pointed out that
fluctuations in the Barnett magnetization will prevent
perfect alignment even if the Barnett timescale is
many orders of magnitude shorter than the timescales
for other processes.
In principle, one could incorporate the Barnett
fluctuations directly in the dynamical simulations.
However, the large disparity between the Barnett timescale
and the timescales associated with other processes makes
such an approach prohibitively expensive.
Here I describe an approximation which eliminates this problem.
The technique developed in this paper will be exploited in
subsequent papers in the series on Davis-Greenstein aligment of
oblate and prolate grains (Roberge \& Lazarian 1997a,b), 
Gold's mechanism (Roberge \& Lazarian 1997c) and superparamagnetic
relaxation (Roberge 1997).

The structure of this paper is as follows: In \S2, the timescales
for relevant dynamical processes are briefly reviewed and the fast
and slow processes are identified.
In \S3, the adiabatic approximation is described and applied
to develop a perturbative solution to the Fokker-Planck equation.
The application of the perturbative approach and its accuracy
are illustrated with a simple example in \S4.
The main results are discussed in \S5 and summarized in \S6.

%
%%%%%%%%%%%%%%%%%%%%%%%%%%%%%%%%%%%%%%%%%%%%%%%%%%%%%%%%%%%%%%%%%%%%%%%%%%%%
% 2. Fast and slow variables
%%%%%%%%%%%%%%%%%%%%%%%%%%%%%%%%%%%%%%%%%%%%%%%%%%%%%%%%%%%%%%%%%%%%%%%%%%%%
%
\section{Fast and slow variables}

Consider the polarization of radiation by an ensemble of aspherical,
partially-aligned dust grains.
Throughout this paper, I will assume for concreteness that the grains
are oblate spheroids; the generalization to other shapes is
straightforward.
The alignment of an ensemble of spheroids is characterized completely by
the ``Rayleigh reduction factor,''
\be
R = \frac{3}{2}\,\left[\,\left<\cos^2\zeta\right>-\frac{1}{3}\,\right]
\label{2_1}
\ee
(Greenberg 1968),
in the sense that $R$ is the only orientation-dependent
quantity that appears in the equations of transfer for the
Stokes parameters.
Here $\zeta$ is the angle between the grain symmetry
axis, \veca, and the magnetic field, \vecB, and the angle brackets
denote averaging over the grain ensemble.

The value of $\zeta$ for a particular grain can be specified by
giving
(i) the orientation of \veca\ relative to
the grain angular momentum, \vecJ\ (see Fig.~1), and
(ii) the orientation of \vecJ\ relative to \vecB\ (see Fig.~2).
With this choice of variables, $R$ depends, in principle,
on the joint distribution of four angles:
$\beta$, $\phi$, $\theta$, and $\psi$.
However, the rapid precessional motions of \veca\ about
\vecJ\ (free precession) and of \vecJ\ about \vecB\ (Larmor precession)
insure that the distributions of $\psi$ and $\phi$ are uniform.
After writing $\cos^2\zeta$ in terms of
$\beta$, $\phi$, $\theta$, and $\psi$ and averaging the resulting
expression over $\psi$ and $\phi$, one finds that
\be
\cos^2\zeta = \frac{1}{2}\,
\left[\,
1-\cos^2\theta-\cos^2\beta+3\,\cos^2\theta\cos^2\beta
\,\right].
\label{2_2}
\ee
The goal of grain alignment theory is to predict
the joint distribution of $\theta$ and $\beta$, and hence $R$,
from a model of the grain dynamics.

The motion of \veca\ with respect to \vecJ\ is determined primarily
by internal relaxation due to the Barnett effect (P79; L94; LR97).
The characteristic timescale for Barnett relaxation in an oblate
spheroid is
\be
\tBar(J) = {\eta^2I_{\parallel}^3 \over VKh^2(h-1)J^2}
\label{2_3}
\ee
(LR97),
where $\eta$ is the magnetogyric ratio of the paramagnetic moments
responsible for the Barnett effect,
\Ipar\ and \Iper\ are the inertias for rotation with \vecJ\ parallel
and perpendicular to the symmetry axis, respectively,
$h\equiv\Ipar/\Iper$, $V$ is the grain volume, and $K$ is
related to the imaginary part of the magnetic susceptibility
at frequency $\omega$ by
\be
K \equiv {{\rm Im}[\chi\left(\omega\right)]\over\omega}.
\label{2_4}
\ee
A typical angular momentum for thermal rotation at the gas
temperature, \Tg, is
\be
J_{\rm therm} \equiv \sqrt{\Ipar k T_{\rm g}}.
\label{2_5}
\ee
Substituting \Jtherm\ into the equation (\ref{2_3}),
one finds that the Barnett time is typically
\be
\tBar(\Jtherm) = 4\times 10^6\,\rho_{{\rm s}0}^2\,T_{{\rm g}1}^{-1}\,
     K_{-13}^{-1}\,b_{-5}^7\,
     {(a/b)\left(1+a^2/b^2\right)^3\over\left(1-a^2/b^2\right)}
     ~~~{\rm s},
\label{2_6}
\ee
where $\rhos$ is the density of the grain material,
$a$ and $b$ are the grain semiaxes parallel and perpendicular to the symmetry
axis, respectively, $b_{-5} \equiv b/\left(10^{-5}\,{\rm cm}\right)$,
and similarly for the scaling of other variables in cgs units.

The motion of \vecJ\ relative to \vecB\ is determined by gas damping
and other external interactions on a timescale which is typically
long compared to the Barnett time.
For example, the characteristic timescale for gas damping
in a cloud of particles with mass $m$ and number density $n$ is
\be
\tgas =
\frac{3I_{\parallel}}{4\sqrt{\pi}nmb^{4}v_{\rm th}\Gpar},
\label{2_7}
\ee
where $\vth\equiv\sqrt{2kT_{\rm g}/m}$ and
$\Gpar$ is a dimensionless coefficient that depends weakly on
the grain shape (see Appendix~A).
For a gas of molecular hydrogen,
\be
\tgas =
7 \times 10^9 \,
\rho_{{\rm s}0}\,
n_{4}^{-1}\,
T_{{\rm g}1}^{-1/2}\,
a_{-5}\,
\Gamma_{\|}\ \ {\rm s}.
\label{2_8}
\ee
Comparing expressions (\ref{2_6}) and (\ref{2_8}), we see that
$\tBar/\tgas \sim 10^{-3}$--$10^{-2}$ typically.
Accordingly, I will refer to $\theta$ and $\psi$ as
the ``fast variables'' and $J$, $\beta$, and $\phi$
as the ``slow variables.''

%
%%%%%%%%%%%%%%%%%%%%%%%%%%%%%%%%%%%%%%%%%%%%%%%%%%%%%%%%%%%%%%%%%%%%%%%%%%%%
% 3. The adiabatic approximation
%%%%%%%%%%%%%%%%%%%%%%%%%%%%%%%%%%%%%%%%%%%%%%%%%%%%%%%%%%%%%%%%%%%%%%%%%%%%
%
\newpage
\section{The adiabatic approximation}

\subsection{The grain alignment problem}

Let \xset\ denote the coordinates of \vecJ\ in the external frame.
It will be useful to take \xset\ to be Cartesian coordinates,
\Jgcart, in some problems and spherical polar coordinates,
\Jgpolar, in others (see Fig.~2).
Let \fjoint\ be the joint distribution function for
$\theta$, $\psi$, and $x$ [with $x$ shorthand for \xset], defined so that
$\fjoint\,dx_1\,dx_2\,dx_3\,d\theta\,d\psi$
is the joint probability of finding the angular momentum in
$dx_1\,dx_2\,dx_3$ and the symmetry axis oriented in
$d\theta\,d\psi$.
By choosing an appropriate set of dimensionless time and
angular momentum units, it is always possible to write
the Fokker-Planck equation for the stationary distribution in the form
\be
\Lintv\,\fjoint + \eps\,\Lextv\,\fjoint =0,
\label{3_1}
\ee
where \Lint\ and \Lext\ are linear differential operators.
The notation $\left(\vint\,|\,\vext\right)$ means that \Lint\ involves
differentiation with respect to $\theta$ but depends on $x$
only as a parameter and similarly for \Lext.
Here $\epsilon$ is the ratio of the timescale for internal
relaxation to the shortest characteristic timescale for
external interactions.
How this all works out for a particular example is demonstrated in \S4.

The operator \Lint\ represents the motion of
$\theta$ due to the combined effects of internal
relaxation and external processes.\footnote{Recall that external
interactions generally alter $\theta$.
For example, a gas-grain collision changes \vecJ\ 
on a timescale that is much shorter than the grain rotation period.
The impulsive motion of \vecJ\ relative to the body axes
generally implies a change in $\theta$ (see Fig.~1).}
It has the form
\be
\Lintv =
-{\partial\over\partial\theta}\,\Atv\,\cdot\ \ 
+\ 
\frac{1}{2}\,{\partial^2\over\partial\theta^2}\,
\Bttv\,\cdot\ \ 
\ + \ \Order(\epsilon),
\label{3_2}
\ee
where \Atv\ and \Bttv\ are the diffusion coefficients for the
internal relaxation mechanism (e.g., Barnett relaxation)
and the unspecified terms of order $\epsilon$ represent the external
interactions.
The diffusion coefficients for internal relaxation
depend on the magnitude of \vecJ\ but not on its orientation in
the external frame, as the notation suggests.
The operator
\be
\Lextv =
-{\partial\over\partial x_m}\,
\Amv\,\cdot\ \ 
+\ 
{1 \over 2}\,{\partial^2\over\partial x_m \partial x_n}\,
\Bmnv\,\cdot
\label{3_3}
\ee
represents the motion of \vecJ\ due to
gas-grain collisions and other external interactions.
The coefficients \Am\ and \Bmn\ generally depend on $\theta$ as
well as $x$.

\subsection{Adiabatic elimination of the fast variable}

We are interested in situations where
$\epsilon \sim 10^{-3}$--$10^{-2}$ typically (see \S2).
A solution of equation (\ref{3_1}) that is accurate to
$\Order(\epsilon)$ should therefore
be satisfactory given the various other
uncertainties in the problem, for example in modeling the grain shape. 
Now if we merely omit all of the terms of order $\epsilon$,
then equation (\ref{3_1}) becomes
\be
-{\partial\over\partial\theta}\,
\left[\,\Atv\,f\,\right]
+
\frac{1}{2}\,{\partial^2\over\partial\theta^2}\,
\left[\,\Bttv\,f\,\right]
=0.
\label{3_4}
\ee
This is the Fokker-Planck equation for
internal relaxation in an isolated grain, all of the external
interactions having been omitted.
Equation (\ref{3_4}) determines the $\theta$ dependence of $f$
but contains no information about the dependence on $J$,
which appears in eq.\ (\ref{3_4}) only as an arbitrary parameter.
Of course, this is appropriate for an isolated grain, whose
angular momentum must be specified as an initial condition.
Alternatively, one may say that equation (\ref{3_4}) predicts the
{\it conditional}\/ probability distribution for $\theta$ given $J$.
Henceforth I will denote the solution of equation (\ref{3_4})
by \fintv\ to emphasize this point.

The solution of equation (\ref{3_4}) does not depend on the
particulars of the internal relaxation mechanism.
In a steady state, the transfer of energy by internal relaxation
between the rotational and vibrational modes of the grain
establishes an equilibrium $\theta$ distribution that depends
only on the laws of thermodynamics and not on the kinetics of
the transfer mechanism (see LR97 for a detailed explanation).
The equilibrium distribution is
\be
\fintv = C(J)\,\snt\,\exp\left(-\xi^2\sin^2\theta\right),
\label{3_5}
\ee
where $C$ is a normalization constant,
\be
\xi^2 \equiv {\left(h-1\right)J^2 \over 2I_{\parallel}kT_s},
\label{3_6}
\ee
and \Ts\ is the temperature of the solid material.
Two related quantities which appear frequently in the following discussion
are the conditional mean value of $\cos^2\theta$,
\be
\chi(J) \equiv
\int_0^{2\pi} d\psi \int_0^{\pi}d\theta\,\cos^2\theta\,\fintv,
\label{3_7}
\ee
and a statistic which measures the alignment of \veca\ with
respect to \vecJ,
\be
Q_X(J) \equiv \frac{3}{2}\left[\,\chi(J)-\frac{1}{3}\,\right].
\label{3_8}
\ee

In summary, expression (\ref{3_5}) gives \fintv\ accurate
to \Ordeps.
This solution was obtained by neglecting the external
interactions, and hence the motion of \vecJ, in computing
the $\theta$ distribution.
This approach assumes, effectively, that the changes in \vecJ\ caused
by the external interactions are so slow that
the $\theta$ distribution relaxes {\it adiabatically}\/ to the solution
that would obtain if the angular momentum coordinates were frozen
at their instantaneous values.
Henceforth I will refer to this assumption as the adiabatic approximation.
Note that the adiabatic approximation is completely analogous
to the Born-Oppenheimer approximation for the wavefunction
of a molecule: in the Born-Oppenheimer approximation,
the motions of the nuclei are assumed to be so slow that the
electronic wavefunction relaxes adiabatically
to the time-independent wavefunction that would obtain if
the nuclear coordinates were frozen at their instantaneous values.

\subsection{Fokker-Planck equation for the slow variables}

Once the conditional distribution is known approximately,
the ``full'' distribution, \fjoint, can be determined as
follows.
Introduce the function \fextv\ defined by
\be
\fjoint \equiv \fintv\,\fextv,
\label{3_9}
\ee
with $J$ regarded as a function of $x$.
Clearly, the rules for computing conditional probabilities
imply that \fext\ is just the distribution function for the
angular momentum coordinates.
The normalization of $f$ and \fint\ together imply that
\be
\int d^3x\,\fextv=1,
\label{3_10}
\ee
consistent with this interpretation.

The dynamical equation for \fext\ follows directly from
equation (\ref{3_1}).
Substitute expression (\ref{3_9}) into equation (\ref{3_1})
and use the fact that \fext\ commutes with \Lint\ to find
\be
\fextv\,\Lintv\,\fintv +
\eps\,\Lextv\,\fintv\,\fextv =0.
\label{3_11}
\ee
Now multiply equation (\ref{3_11}) 
by $d\theta\,d\psi$ and integrate over
$\theta$ and $\psi$.
The first term on the left side vanishes upon integration,
leaving
\be
\int_0^{2\pi} d\psi
\int_0^{\pi} d\theta\,\Lextv\,\fintv\,\fextv =0.
\label{3_12}
\ee
After substituting expression (\ref{3_3}) for \Lextv\ 
and carrying out the integration, one finds that
equation (\ref{3_12}) is equivalent to
\be
-{\partial\over\partial x_m}\,
\left[\,\Abmv\fextv\,\right] 
+
{1 \over 2}\,{\partial^2\over\partial x_m \partial x_n}\,
\left[\,\Bbmnv\fextv\,\right] =0,
\label{3_13}
\ee
where
\be
\Abmv \equiv \int_0^{2\pi}d\psi\int_0^{\pi}d\theta\ 
\Amv\,\fintv
\label{3_14}
\ee
and
\be
\Bbmnv \equiv \int_0^{2\pi}d\psi\int_0^{\pi}d\theta\ 
\Bmnv\,\fintv.
\label{3_15}
\ee
\noindent
Expression  (\ref{3_13}) is the main result of this paper.
It is a Fokker-Planck equation for the angular momentum coordinates
wherein the fast variable, $\theta$, has been eliminated by the
averaging in equations (\ref{3_14}) and (\ref{3_15}).
Equations (\ref{3_14}) and (\ref{3_15}) say 
that the effective diffusion coefficients for the angular momentum
coordinates, \Abm\ and \Bbmn, are obtained by averaging
the $\theta$-dependent coefficients \Am\ and \Bmn\ in the
obvious way over the equilibrium $\theta$ distribution.
The analysis in this section establishes
that the averaging procedure is accurate to \Ordeps.

\subsection{Calculating the Rayleigh reduction factor}

The calculation of the Rayleigh reduction factor simplifies
in the adiabatic approximation.
After multiplying expression (\ref{2_2}) by the joint distribution,
expression (\ref{3_9}), and carrying out the integration over
$\theta$ analytically, one finds that
\be
R = \int\ d^3x\,\fextv\,\QX(J)\,Q_J(\beta),
\label{3_16}
\ee
where $J$ and $\beta$ are regarded as functions of $x$,
$\QX(J)$ is a known function of $J$
(see eqs.\ [\ref{3_5}]--[\ref{3_8}]), and
\be
Q_J\left(\beta\right) \equiv
\frac{3}{2}\left(\cos^2\beta-\frac{1}{3}\right).
\label{3_17}
\ee

%
%%%%%%%%%%%%%%%%%%%%%%%%%%%%%%%%%%%%%%%%%%%%%%%%%%%%%%%%%%%%%%%%%%%%%%%%%%%%
% 4. An illustrative example
%%%%%%%%%%%%%%%%%%%%%%%%%%%%%%%%%%%%%%%%%%%%%%%%%%%%%%%%%%%%%%%%%%%%%%%%%%%%
%
\section{An illustrative example}

As a simple application of the approach described in \S3,
consider the problem of finding \fext\ when the
internal relaxation is produced by the Barnett effect and the
external interactions are due entirely to gas damping.
This problem was solved numerically in a study of Barnett relaxation by
Lazarian \& Roberge (1997).
Here I show that all the results of practical interest can be
obtained analytically in the adiabatic approximation.

\subsection{Fokker-Planck equation for the angular momentum distribution}

Adopt a system of dimensionless variables with angular momentum
measured in units of \Jtherm\ and time in units of \tBarJt\
(see eqs.\ [\ref{2_3}] and [\ref{2_5}]).
In this system, the diffusion coefficients for Barnett
relaxation (see \S A.1) are all of order unity and those
for gas damping (see \S A.2) are expressions of order unity times
\be
\eps \equiv {\tBar\left(J_{\rm therm}\right) \over t_{\rm gas} }
\label{4_1}
\ee
(compare
eq.\ [\ref{A_3}] with eqs.\ [\ref{A_14}]--[\ref{A_16}]
and eq.\ [\ref{A_4}] with eqs.\ [\ref{A_18}]--[\ref{A_20}]).
This verifies that the dimensionless Fokker-Planck equation
conforms to expression (\ref{3_1}) and hence
that the analysis of \S3 applies.

To proceed, we wish to solve equation (\ref{3_13}) with the
coefficients \Abm\ and \Bbmn\ appropriate for gas damping.
To exploit the absence of a preferred direction in the
external frame, it is advantageous to adopt spherical
polar angular momentum coordinates, \Jgpolar.
After transforming\footnote{The transformation of the
Fokker-Planck equation from one coordinate
system to another is described in standard texts
on statistical mechanics (e.g., see Risken 1984).
The procedure is straightforward but tedious and I will
only give the relevant results here.}
the diffusion coefficients in Appendix~A\footnote{Modulo the
factor \eps; compare eqs.\ (\ref{3_1}) and (\ref{3_3}).}
from Cartesian to polar coordinates, equation (\ref{3_13})
becomes\footnote{I have actually taken a slight liberty in writing
equation (\ref{4_2}). The linear coefficients
$\bar{A}_{\beta}$ and $\bar{A}_{\phi}$ are zero, as equation (\ref{4_2})
implies, but the quadratic coefficients
$\bar{B}_{\beta\beta}$, $\bar{B}_{\phi\phi}$, etc are not.
However, there is no harm in omitting the nonzero quadratic terms:
they represent isotropic diffusion in the angular coordinates and,
since the linear terms are zero, they merely insure
that \fext\ is isotropic.
There is no need to include the quadratic terms explicitly 
if we assume a priori that \fext\ is isotropic.}
\be
-{d\over dJ}\left\{\,
\AbJ\,\fext-\frac{1}{2}\,{d\over dJ}\left[\BbJJ\,\fext\right]
\right\}=0,
\label{4_2}
\ee
where the ``polar'' diffusion coefficients,
\be
\AbJ = -k\,J +
\frac{1}{2J}\,\left[2\Bper+\sigma\,\left(\Bpar-\Bper\right)\right]
\label{4_3}
\ee
and
\be
\BbJJ = \sigma\,\Bper\,+\chi\,\Bpar,
\label{4_4}
\ee
depend on $J$ but are independent of $\beta$ and $\phi$.
Here
\be
k \equiv h\gamma\,\sigma+\chi
\label{4_5}
\ee
and
\be
\sigma \equiv 1-\chi
\label{4_6}
\ee
are functions of $J$, $\gamma$ depends only on the grain shape
(see eq.\ [\ref{A_17}]), and the dimensionless quantities
\be
\Bper = \gamma\left(1+\Ts/\Tg\right)
\label{4_7}
\ee
and
\be
\Bpar = \left(1+\Ts/\Tg\right)
\label{4_8}
\ee
depend on the grain shape and dust-to-gas temperature ratio.

\subsection{Closed-form solution}

Equation (\ref{4_2}) is a second-order differential equation for
$\fext$.
It can be simplified by noting that the quantity in curly brackets
is the probability current along the \vecJ\ direction.
Setting the current to zero yields
linear, first-order differential equation for \fext,
\be
\AbJ\,\fext-\frac{1}{2}\,{d\over dJ}\left[\BbJJ\,\fext\right]=0.
\label{4_9}
\ee
Noting that we have defined \fext\ to be the probability per unit
coordinate interval, $dJ$, we see that the appropriate boundary
condition is
\be
\lim_{J\rightarrow 0} \,\fext = 0.
\label{4_10}
\ee

One can easily verify that the solution of equation (\ref{4_9})
subject to boundary condition (\ref{4_10}) is
\be
\fext(J,\beta) = C\,J^2\,\sin\beta\,
\frac{\exp\left[F\left(J\right)\right]}{\bar{B}_{JJ}},
\label{4_11}
\ee
where $C$ is a normalization constant,
the factor of $\sin\beta$ is required to insure that \fext\ is isotropic,
\be
F\left(J\right) \equiv \int_{J_0}^J\ 
\lambda\left(J^{\prime}\right)\,dJ^{\prime},
\label{4_12}
\ee
$J_0$ is a an arbitrary constant, and
\be
\lambda(J) \equiv
{2\bar{A}_J\over \bar{B}_{JJ}}-{2\over J}.
\label{4_13}
\ee
This completes the solution.
The expressions for $\bar{A}_J$ and $\bar{B}_{JJ}$ imply
that \fext\ depends on just two dimensionless parameters,
$a/b$ and \TsTg.

\subsection{Spherical grains ({\boldmath$a/b=1$})}

A check on the closed-form solution is provided by the special
case of spherical grains.
For spheres, the diffusion coefficients for gas damping
are independent of $\theta$ (see \S A.2), so Barnett relaxation can
be ignored.
The angular momentum distribution of spherical grains subject
only to gas damping is Maxwellian with an effective temperature
$T_{\rm av}=\frac{1}{2}\left(\Ts+\Tg\right)$.
In dimensionless units, the Maxwellian distribution for spheres is
\be
\fext(J,\beta) = C\,J^2\,\sin\beta\,
     \exp\left[-{J^2\over\left(1+T_{\rm s}/T_{\rm g}\right)}\right].
\label{4_14}
\ee

To see whether the closed-form solution reproduces expression
(\ref{4_14}),
note that $\gamma=h=k=1$ and $\Bpar=\Bper=\Tfac$ for spheres.
Taking $J_0=0$ in equation (\ref{4_12}), we find that
\be
\lambda = -{2J \over \left(1+T_{\rm s}/T_{\rm g}\right)}
\label{4_15}
\ee
and
\be
F\left(J\right) = -J^2/\Tfac.
\label{4_16}
\ee
This proves that the closed-form solution is exact for
spheres.

\subsection{Equal gas and dust temperatures ({\boldmath$\Ts=\Tg$})}

If $\Ts=\Tg$, then all of the temperatures in the system are
equal and the angular momentum distribution must be
Maxwellian even if the grains are not spherical.
It is possible to prove that the closed-form solution reduces
to the Maxwellian distribution when $\Ts=\Tg$ but the derivation
is tedious.
As a simpler but weaker test, one can compare the statistics
of the closed-form solution to those of the Maxwellian.
The isotropy of \fext\ insures that $R=0$, as required.
Consider next the statistic
\be
\QX \equiv \int_0^{\pi}\ d\beta\,\int_0^{\infty}\ dJ\ \QX(J)\,
\fext(J,\beta),
\label{4_17}
\ee
which measures the alignment of \veca\ with respect to \vecJ\
(see eq.\ [\ref{3_8}]).
For a Maxwellian distribution, $\QX=\QXMax$, where
\be
\QXMax =
{3 \over 2 \left(1-h^{-1}\right)}\,
\left[\,1-{1\over \sqrt{h-1}}\,\sin^{-1}\left(1-h^{-1}\right)\,\right]
-
\frac{1}{2}
\label{4_18}
\ee
(Jones \& Spitzer 1967; LR97) and $h=2/\left(1+a^2/b^2\right)$
for a homogeneous spheroid.
Figure~3 is a comparison between the values of \QXMax\
given by expression (\ref{4_18}) (solid curve) and the \QX\ values
predicted by the closed-form solution (open circles).

\subsection{The general case}

The general solution is shown in Figure~4, where \QX\ 
has been plotted vs.\ \TsTg\ for selected values of $a/b$.
The solid curves are the predictions of the adiabatic approximation;
they were calculated by evaluating the integrals
in expressions (\ref{4_12}) and (\ref{4_17}) numerically,
but otherwise they are analytic results.
The symbols in Figure~4 are the results of a numerical
calculation which did not use the abiatic approximation (LR97).
It took $>90$ CPU hours on an RS/6000 workstation to
compute these numerical solutions.
The symbols in Figure~4 correspond to the case $\epsilon=0.04$,
that is, to a case which is close to the limit,
$\epsilon \rightarrow 0$, where the adiabatic approximation
becomes exact.
Notice that the discrepancies between the analytic and numerical
solutions in Figure~4 are comparable to the fluctuations in the
numerical results.

%
%%%%%%%%%%%%%%%%%%%%%%%%%%%%%%%%%%%%%%%%%%%%%%%%%%%%%%%%%%%%%%%%%%%%%%%%%%%%
% 5. Discussion
%%%%%%%%%%%%%%%%%%%%%%%%%%%%%%%%%%%%%%%%%%%%%%%%%%%%%%%%%%%%%%%%%%%%%%%%%%%%
%
\section{Discussion}

This paper shows that, whenever the timescale for internal relaxation
is small compared to the timescales for external processes, one can
solve the grain alignment problem without explicitly considering
the rapid ``internal'' motions of the grain axes relative to \vecJ.
In this regime, the problem reduces to solving a Fokker-Planck equation
for the angular momentum coordinates.
This approach has significant practical advantages.
Simulation techniques for solving the Fokker-Planck
equation (e.g., RDGF93) must follow the dynamics with a time step
that is typically
$\sim 10^{-3}$ times the smallest timescale of the motion
and average the motion over timescales $\sim 10^5$ times
the longest characteristic time to calulate $R$ accurate to
a few percent.
According to \S2, the characteristic timescales for the internal
and external variables differ by a factor of
$\epsilon^{-1} \sim 10^2$--$10^3$, so a straightforward approach that
simulated the internal and external motions directly
would require $\sim 10^{11}$ time steps to compute each
$R$ value.
The adiabatic elimination of the internal variables
reduces the computational burden by a factor of $\epsilon$
and the solution of equation (\ref{3_13}) by simulation techniques,
though computationally intensive, is feasible.
In the next papers in the series, we follow this approach
to predict the efficiency of the Davis-Greenstein mechanism
(Roberge \& Lazarian 1997a,b), Gold's mechanism (Roberge \& Lazarian 1997c)
and superparamagnetic alignment (Roberge 1997).

%
%%%%%%%%%%%%%%%%%%%%%%%%%%%%%%%%%%%%%%%%%%%%%%%%%%%%%%%%%%%%%%%%%%%%%%%%%%%%
% 6. Summary
%%%%%%%%%%%%%%%%%%%%%%%%%%%%%%%%%%%%%%%%%%%%%%%%%%%%%%%%%%%%%%%%%%%%%%%%%%%%
%
\clearpage
\section{Summary}

The principal results of this paper are as follows:

\begin{enumerate}

\item
The adiabatic approximation greatly simplifies grain
alignment calculations in regimes where large disparities
exist between the dynamical timescales associated with different processes.
The adiabatic approximation prescribes an analytic solution
for the distribution function of the fast variables.
The slow variables are described by a Fokker-Planck equation
wherein the fast variables have been eliminated by a
simple averaging procedure.
\item
The adiabatic approximation is accurate to \Ordeps, where \eps\ is the
ratio of the dynamical timescales for the fast and slow variables.
\item
The adiabatic approximation provides an efficient technique for
incorporating the physics of internal relaxation in studies
of grain alignment.
In a typical application, the adiabatic approximation
reduces the computational effort required to obtain a
solution by a factor $\sim 10^2$--$10^3$.
The error introduced by the approximation is typically
only a few parts in a thousand.
\item
The approach described in this paper has been used to
find an analytical solution to a problem solved earlier
using numerical techniques (LR97).
Comparing the analytic and numerical results illustrates
the simplicity and accuracy of the method advocated here.
It also verifies the accuracy of the numerical calculations.

\end{enumerate}

\acknowledgements
This work was partially supported by NASA grant NAGW-3001.
I thank Robert Lupton for stimulating my interest in
asymptotic solutions and Joe Haus for helpful advice on
the adiabatic elimination technique.
It is a pleasure to acknowledge Dave Messinger for helpful comments
and Alex Lazarian for a careful reading of the original
manuscript.

%
%%%%%%%%%%%%%%%%%%%%%%%%%%%%%%%%%%%%%%%%%%%%%%%%%%%%%%%%%%%%%%%%%%%%%%%%%%%%
% REFERENCES
%%%%%%%%%%%%%%%%%%%%%%%%%%%%%%%%%%%%%%%%%%%%%%%%%%%%%%%%%%%%%%%%%%%%%%%%%%%%
%

%
%%%%%%%%%%%%%%%%%%%%%%%%%%%%%%%%%%%%%%%%%%%%%%%%%%%%%%%%%%%%%%%%%%%%%%%%%%%%
% A. Diffusion coefficients
%%%%%%%%%%%%%%%%%%%%%%%%%%%%%%%%%%%%%%%%%%%%%%%%%%%%%%%%%%%%%%%%%%%%%%%%%%%%
%
\appendix
\section{Diffusion coefficients}

\subsection{Barnett relaxation}

The diffusion coefficients for Barnett relaxation were derived in LR97.
In ordinary (dimensional) units,
they are\footnote{The careful reader may notice that expressions
(\ref{A_1}) and (\ref{A_2}) differ from the analogous
expressions in LR97 by factors of $J$ and $J^2$, respectively.
This is appropriate: LR97 described Barnett relaxation in a coordinate
system fixed to the grain, where the Barnett effect appears
to move \vecJ. Appropriately, LR97 identified $A_{\theta}$ with the
fictitious mean torque required to produce this motion.
Here I describe the dynamics in an inertial frame, where
\vecJ\ and $\theta$ are {\it independent}\/ variables.
In this view, $A_{\theta}$ is the mean rate of change of $\theta$,
not a torque.}
\be
\hat{A}_{\theta} = -\,{\sin\theta\cos\theta \over t_{\rm Bar}(\hat{J})}
\label{A_1}
\ee
and
\be
\hat{B}_{\theta\theta} =
\frac{1}{t_{\rm Bar}(\hat{J})\sin\theta}\,\left[\,
\exp(\xi^2\sin^2\theta)
\int^{1}_{\sin^2\theta}\sqrt{y}\exp(-\xi^2 y){\rm d}y +
\exp(-\xi^2\cos^2\theta)\,\right],
\label{A_2}
\ee
where the hats denote dimensional quantities to avoid confusion
with their dimensionless counterparts.
In a system of units where
$\tBar\left(\Jtherm\right)$ is the unit of time and
$\Jtherm$ is the unit of angular momentum, the dimensionless
diffusion coefficients are
\be
A_{\theta\theta} = -J^2\,\sin\theta\cos\theta
\label{A_3}
\ee
and
\be
B_{\theta\theta} =
\frac{J^2}{\sin\theta}\,
\left[\,
\exp(\xi^2\sin^2\theta)\,
\int^{1}_{\sin^2\theta}\sqrt{y}\exp(-\xi^2 y){\rm d}y +
\exp(-\xi^2\cos^2\theta)\,\right],
\label{A_4}
\ee
respectively.

\subsection{Gas damping}

The diffusion coefficients for gas damping were also derived in LR97.
Relative to the basis of the internal frame, \bfbasis\ (see Fig.~1), the
mean torque has Cartesian components
\be
\hat{A}^b_x =
-\frac{4\sqrt{\pi}}{3I_{\perp}}nmb^{4}v_{\rm th}\Gamma_{\bot}
\hat{J}^b_x,
\label{A_5}
\ee
\be
\hat{A}^b_y =
-\frac{4\sqrt{\pi}}{3I_{\perp}}nmb^{4}v_{\rm th}\Gamma_{\bot}
\hat{J}^b_y,
\label{A_6}
\ee
and
\be
\hat{A}^b_z =
-\frac{4\sqrt{\pi}}{3I_{\parallel}}nmb^{4}v_{\rm th}\Gamma_{\|}
\hat{J}^b_z.
\label{A_7}
\ee
The dimensionless factors \Gpar\ and \Gper\ are functions of
the eccentricity, $e\equiv\sqrt{1-a^2/b^2}$, with
\be
\Gamma_{\parallel}(e) =
{3 \over 16} \, \left\{\
3+4(1-e^2)g(e)-e^{-2}\left[1-(1-e^2)^2g(e)\right]
\right\}
\label{A_8}
\ee
and
\be
\Gamma_{\perp}(e) =
{3 \over 32} \, \left\{\
7-e^2+(1-e^2)^2g(e)+
(1-2e^2)\left[1+e^{-2}\left[1-(1-e^2)^2g(e)\right]\right]\right\},
\label{A_9}
\ee
respectively, and
\be
g(e) \equiv {1 \over 2e} \ln\left({1+e \over 1-e}\right).
\label{A_10}
\ee
The diffusion tensor for gas damping is diagonal in the
internal frame with components
\be
\hat{B}^b_{xx}=
\frac{2\sqrt{\pi}}{3}nmb^{4}v_{\rm th}^{3}\Gamma_{\bot}(e)
\left(1+\frac{T_{\rm s}}{T_{\rm g}}\right),
\label{A_11}
\ee
\be
\hat{B}^b_{yy}=B^b_{xx},
\label{A_12}
\ee
and
\be
\hat{B}^b_{zz}=
\frac{2\sqrt{\pi}}{3}nmb^{4}v_{\rm th}^{3}\Gamma_{\|}(e)
\left(1+\frac{T_{\rm s}}{T_{\rm g}}\right).
\label{A_13}
\ee

To transform to dimensionless variables, multiply
the mean torque by
$\tBar\left(\Jtherm\right)/\Jtherm$
and the diffusion tensor by $J_{\rm therm}^2/\tBar\left(\Jtherm\right)$.
The dimensionless mean torque is
\be
A^b_x = -\eps\,h\,\gamma\,J^b_x,
\label{A_14}
\ee
\be
A^b_y = -\eps\,h\,\gamma\,J^b_y,
\label{A_15}
\ee
and
\be
A^b_z = -\eps\,J^b_z,
\label{A_16}
\ee
where
\be
\gamma \equiv \Gper/\Gpar.
\label{A_17}
\ee
The nonzero components of the dimensionless diffusion tensor are
\be
B^b_{xx} = \eps\,\gamma\,\left(1+T_{\rm s}/T_{\rm g}\right)
\equiv \eps\,\Bper,
\label{A_18}
\ee
\be
B^b_{yy} = \eps\,\Bper
\label{A_19}
\ee
and
\be
B^b_{zz} = \eps\,\left(1+T_{\rm s}/T_{\rm g}\right)
\equiv \eps\,\Bpar.
\label{A_20}
\ee

%
%%%%%%%%%%%%%%%%%%%%%%%%%%%%%%%%%%%%%%%%%%%%%%%%%%%%%%%%%%%%%%%%%%%%%%%%%%%%
% Figure Captions
%%%%%%%%%%%%%%%%%%%%%%%%%%%%%%%%%%%%%%%%%%%%%%%%%%%%%%%%%%%%%%%%%%%%%%%%%%%%
%
\newpage
\centerline{\bf FIGURE CAPTIONS}

\noindent
{\bf Fig.~1} ---
Angles $\theta$ and $\psi$ specify the orientation of
the symmetry axis, \veca, with respect to the angular momentum,
\vecJ.
The ``internal frame'' attached to the grain has basis
\bfbasis, with \bfzh\ parallel to \veca\ and the other basis
vectors oriented as shown.

\bigskip
\noindent
{\bf Fig.~2} ---
Angles $\beta$ and $\phi$ specify the orientation of
\vecJ\ with respect to the magnetic field, \vecB.
The ``external frame,'' \gfbasis, has \gfzh\ directed
along \vecB\ and the remaining basis vectors
oriented as shown.

\bigskip
\noindent
{\bf Fig.~3} ---
Values of the statistic \QX\ for spherical grains plotted
vs.\ the axis ratio.
Solid curve: exact solution (eq.\ [\ref{4_18}]).
Symbols: predictions of the adiabatic approximation.

\bigskip
\noindent
{\bf Fig.~4} ---
Values of \QX\ for arbitrary grain shapes and dust-to-gas
temperature ratios.
Results are shown for selected values of the axis ratio, $a/b$,
with $a/b=0.1$ (top curve, open circles), $a/b=0.5$
(middle curve, filled circles), and $a/b=0.9$
(bottom curve, triangles).
Solid curves: analytic results obtained with the adiabatic approximation.
Symbols: numerical results from Lazarian \& Roberge (1997), obtained
for the case $\epsilon=0.04$ without using the adiabatic approximation.

%
%%%%%%%%%%%%%%%%%%%%%%%%%%%%%%%%%%%%%%%%%%%%%%%%%%%%%%%%%%%%%%%%%%%%%%%%%%%%
% Postscript figures --- for preprints only
%%%%%%%%%%%%%%%%%%%%%%%%%%%%%%%%%%%%%%%%%%%%%%%%%%%%%%%%%%%%%%%%%%%%%%%%%%%%
%

%
%%%%%%%%%%%%%%%%%%%%%%%%%%%%%%%%%%%%%%%%%%%%%%%%%%%%%%%%%%%%%%%%%%%%%%%%%%%%
% Figure 1
%%%%%%%%%%%%%%%%%%%%%%%%%%%%%%%%%%%%%%%%%%%%%%%%%%%%%%%%%%%%%%%%%%%%%%%%%%%%
%
\clearpage
\begin{figure}
\begin{picture}(450,550)
\includegraphics{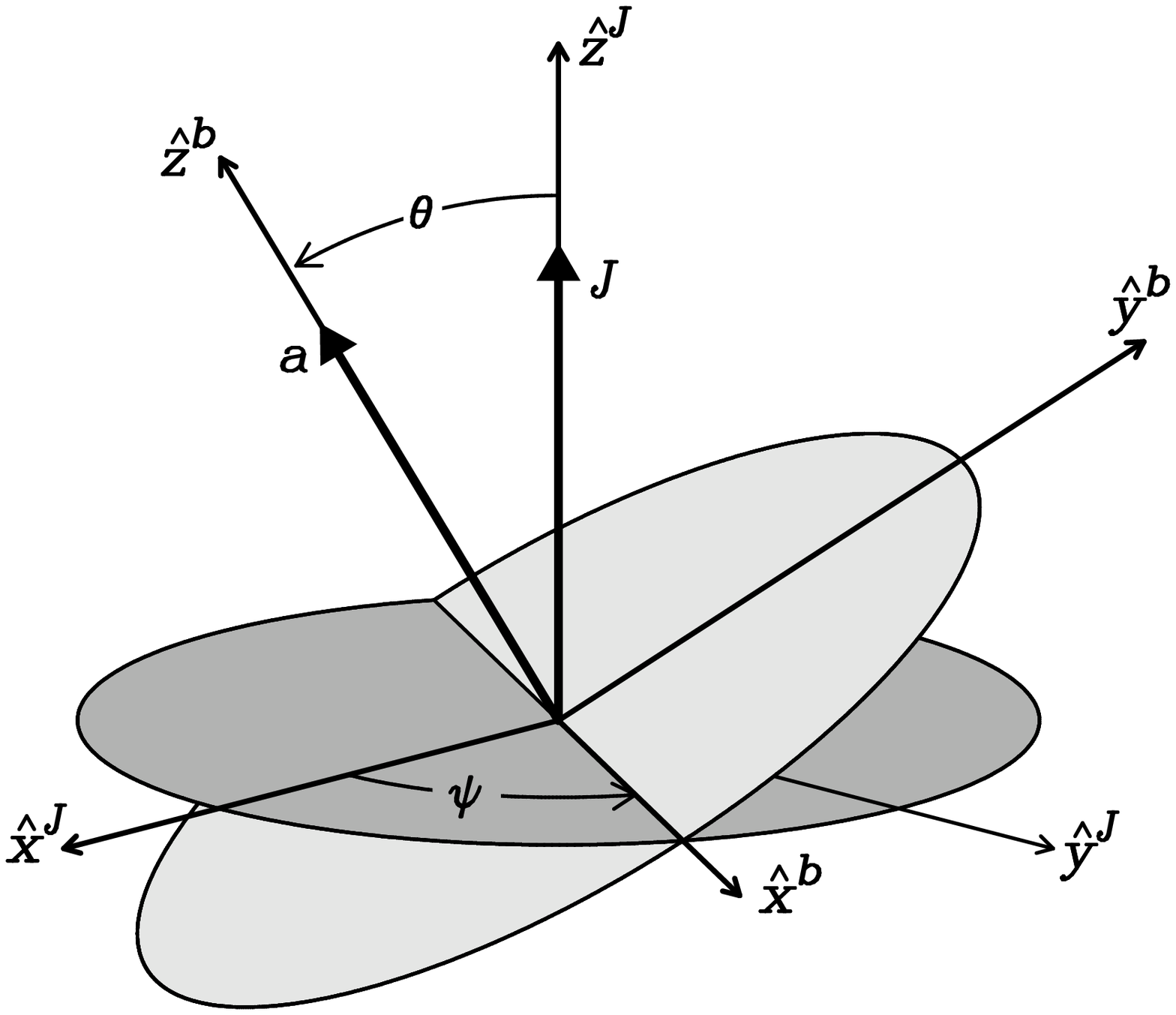}
\smallskip
\centerline{\large\bf Figure\ 1}
\end{picture}
\end{figure}

%
%%%%%%%%%%%%%%%%%%%%%%%%%%%%%%%%%%%%%%%%%%%%%%%%%%%%%%%%%%%%%%%%%%%%%%%%%%%%
% Figure 2
%%%%%%%%%%%%%%%%%%%%%%%%%%%%%%%%%%%%%%%%%%%%%%%%%%%%%%%%%%%%%%%%%%%%%%%%%%%%
%
\clearpage
\begin{figure}
\begin{picture}(450,550)
\includegraphics{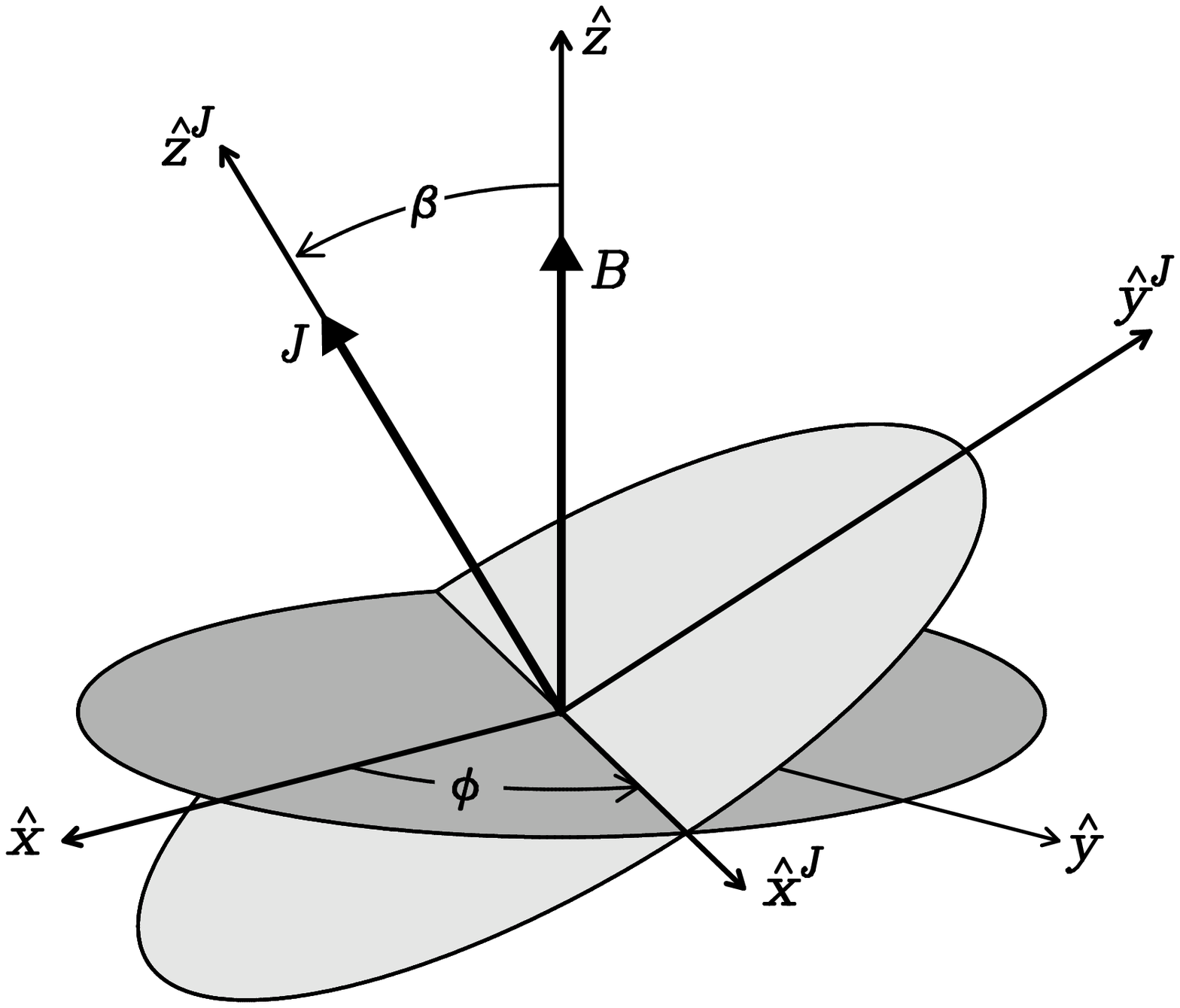}
\smallskip
\centerline{\large\bf Figure\ 2}
\end{picture}
\end{figure}

%
%%%%%%%%%%%%%%%%%%%%%%%%%%%%%%%%%%%%%%%%%%%%%%%%%%%%%%%%%%%%%%%%%%%%%%%%%%%%
% Figure 3
%%%%%%%%%%%%%%%%%%%%%%%%%%%%%%%%%%%%%%%%%%%%%%%%%%%%%%%%%%%%%%%%%%%%%%%%%%%%
%
\clearpage
\begin{figure}
\begin{picture}(450,550)
\includegraphics{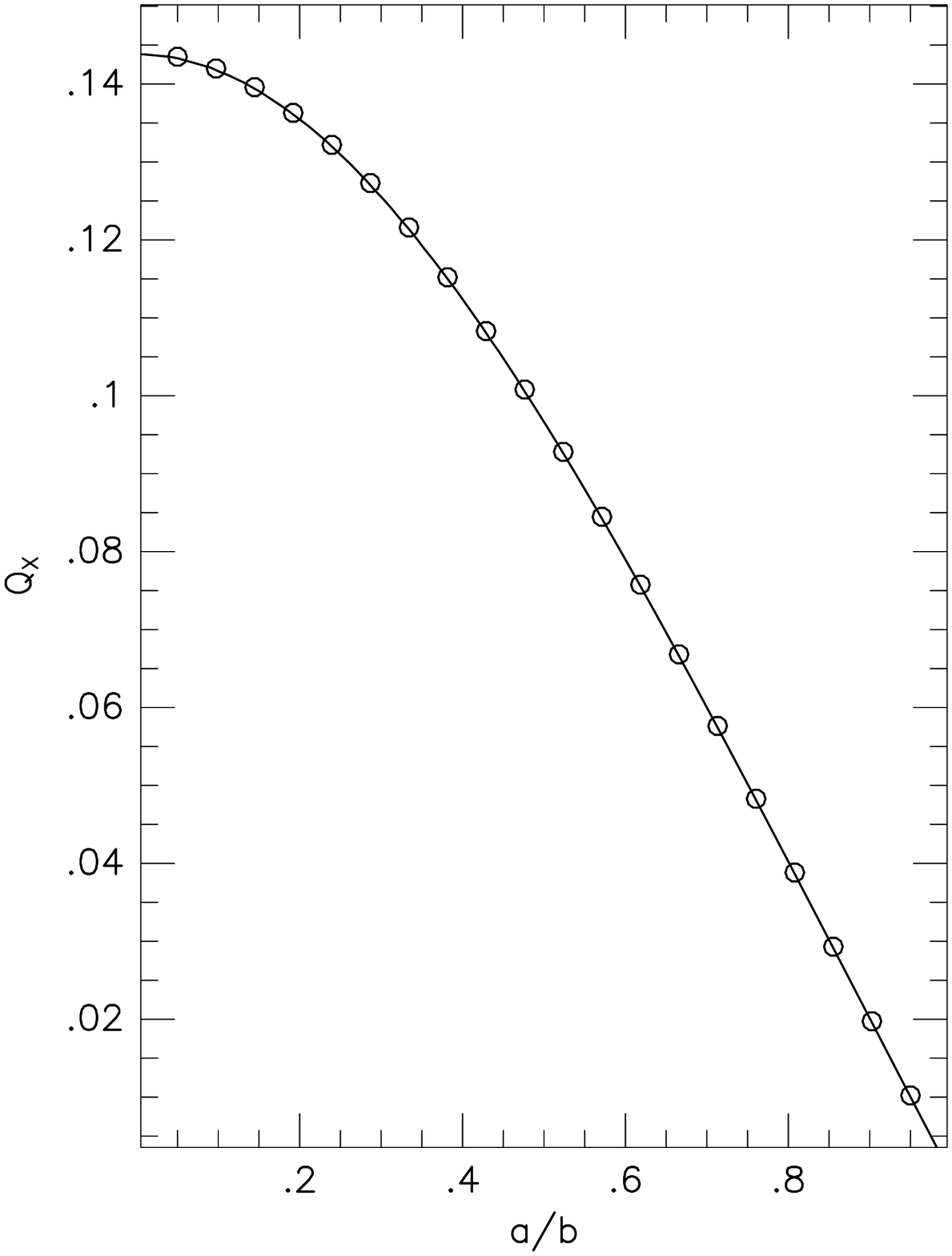}
\smallskip
\centerline{\large\bf Figure\ 3}
\end{picture}
\end{figure}

%
%%%%%%%%%%%%%%%%%%%%%%%%%%%%%%%%%%%%%%%%%%%%%%%%%%%%%%%%%%%%%%%%%%%%%%%%%%%%
% Figure 4
%%%%%%%%%%%%%%%%%%%%%%%%%%%%%%%%%%%%%%%%%%%%%%%%%%%%%%%%%%%%%%%%%%%%%%%%%%%%
%
\clearpage
\begin{figure}
\begin{picture}(450,550)
\includegraphics{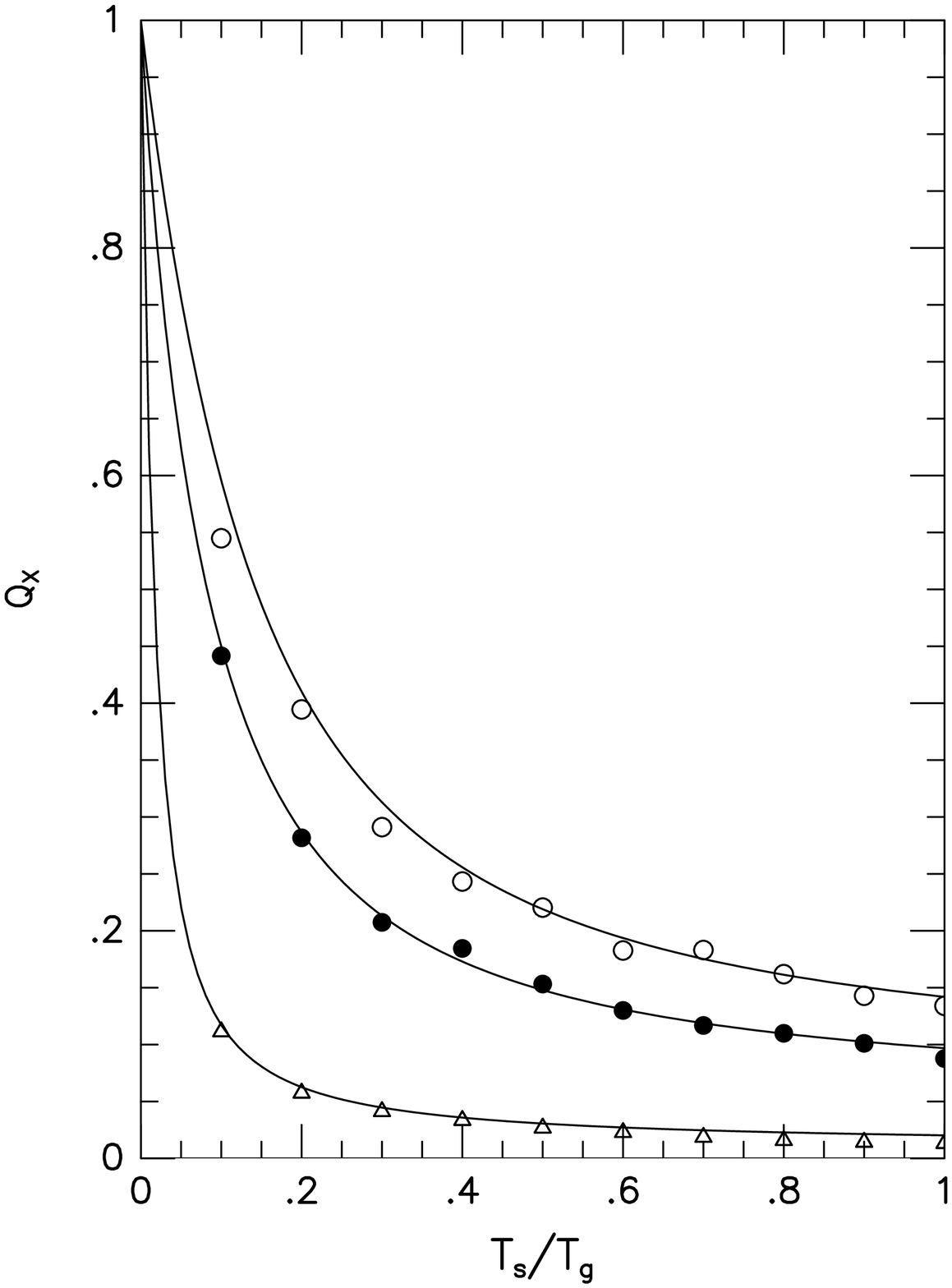}
\smallskip
\centerline{\large\bf Figure\ 4}
\end{picture}
\end{figure}

%
%%%%%%%%%%%%%%%%%%%%%%%%%%%%%%%%%%%%%%%%%%%%%%%%%%%%%%%%%%%%%%%%%%%%%%%%%%%%
% All done
%%%%%%%%%%%%%%%%%%%%%%%%%%%%%%%%%%%%%%%%%%%%%%%%%%%%%%%%%%%%%%%%%%%%%%%%%%%%
%
\end{document}